\documentstyle[prb,aps,twocolumn]{revtex}
\begin{document}
\draft
\title{
Low-energy states for correlated-electron models
in the strong-coupling limit
}
\author{Andreas Giesekus and Uwe Brandt}
\address{Institut f\"ur Physik, Universit\"at Dortmund\\
         D-44221 Dortmund\\
         Germany}
\maketitle
\begin{abstract}
We study a class of exactly solvable models for strongly correlated
electrons, defined on a set of $N$ cells, and with infinite on-site
repulsion on part of the sites of each cell. For $2N$ or more electrons
the exact ground state is known. We construct a tractable
$(2N-1)$-particle state which becomes asymptotically degenerate to
the ground state in the thermodynamic limit for one special
$D$-dimensional model. For other models, that state may be used to
calculate variational upper bounds on the ground-state energy. For a
Hubbard chain with three sites per unit cell, the analytical bound
is compared to numerically exact results.
\end{abstract}

\narrowtext
%

%
\section{INTRODUCTION}
\label{sec:1}

The theoretical description of electronic correlations
in solids is still a challenge. Even one of the
simplest models for correlated electrons, the Hubbard
model,\cite{hubbard} exhibits enormous mathematical
complexity. Its analysis is far from being complete.

Recently, a new class of models for tight-binding electrons
with infinitely strong on-site repulsion has been discovered.
For particle numbers of two or more particles per unit cell,
one can construct the ground states of these
models.\cite{brandt-giesekus} These ground states are
spin-singlets and have the structure of
resonating-valence-bond states\cite{tasaki1} for certain
cases.

The class of models which are solvable by this method
has been generalized by several
authors.\cite{tasaki1,mielke,strack,bares-lee,tasaki2}
One-dimensional models of this class\cite{strack} allowed
for the calculation of equal-time correlation
functions\cite{bares-lee} using a transfer-matrix
technique. All equilibrium correlations studied
so far decay exponentially. Dynamical correlations
obtained numerically for the one-dimensional Hubbard
chain indicate dispersionless excitation spectra for
an electron density of two per cell.\cite{giesekus} The
propagation of one hole, however, shows dispersive
delocalized features.

The regime of particle numbers below two per unit cell
has not been accessible by analytical methods so far. The
main result of this paper is an analytical upper bound on
the ground state energy of systems containing $2N-1$ particles
($N$ denotes the number of cells). The trial state and the
corresponding upper bound on the ground-state energy are
discussed for two examples. The first example is a hypercubic
Hubbard model as introduced in Ref.\onlinecite{brandt-giesekus}.
It is shown that the trial state becomes an asymptotically
exact eigenstate of this particular $(2N-1)$ -particle system
in the thermodynamic limit. For the second system, a linear
Hubbard chain, the upper bound on the ground-state energy may
be calculated analytically as well. A comparison to exact
numerical results\cite{giesekus} is presented.

\section{The class of solvable Model Hamiltonians}
\label{sec:2}

In this section we recall the most general description of the
class of solvable models.\cite{tasaki2} This class contains
Hubbard, Anderson and Emery models in the limit of infinitely
strong interaction ($U=\infty$). The solution presented below
requires certain lattice structures. (In the following,
the term {\em graph} is used, because {\em lattice} implies
translational invariance which needs not be assumed.)

The models are defined on graphs, where each of the $N$ vertices
contains a {\em cell} of sites. (The cells need not be identical
either.) The cells are labeled by an index $i=1,\ldots N$. Let
the set of all sites within the cell $i$ be called ${\cal N}_i$.
(It is mentioned in passing that ${\cal N}_i$ is {\em not} a unit
cell if the graph is translationally invariant, because neighboring
cells may share some sites.)
The sites (or electronic orbitals) within a cell $i$ are labeled
by an index $\alpha$. Some of those orbitals (not necessarily all)
may carry an infinitely strong on-site repulsion, e.\,g.\ a cell
may contain ``$d$-sites'' which may be empty or occupied by one
particle only (with spin up or down) and ``$p$-sites'' which can
be occupied by up to two particles. The set of $d$-sites is called
${\cal U}_{i}$. The repulsion is incorporated into the model by a
projection operator
\begin{equation}
\label{eq:1}
P \quad = \quad \prod_{i}
                      \prod_{\alpha \in {\cal U}_i}
                      \big(
            1 - n_{i,\alpha,\uparrow}^{}n_{i,\alpha,\downarrow}^{}
                      \big)
\end{equation}
which strictly excludes double occupancy on all sites
$i,\alpha \in {\cal U}_i$, where
$n_{i,\alpha,\sigma}^{}=
c_{i,\alpha,\sigma}^{\dagger}
c_{i,\alpha,\sigma}^{}$. The Fermi operators
$c_{i,\alpha,\sigma}^{\dagger}$
($c_{i,\alpha,\sigma}^{}$) denote creation (annihilation) operators
and are defined as usual. The connection between cells is established
by common sites. We denote the set of common sites by the intersection
${\cal C}_{ij}={\cal N}_i \cap {\cal N}_j$. Examples for some models
are given in Ref.\onlinecite{brandt-giesekus,tasaki1,tasaki2}.
Figure \ref{fig:1} illustrates the two cases discussed below.

In order to construct the ground state of the models defined
above consider the linear combination of Fermi operators
\begin{equation}
\label{eq:2}
{\Psi}_{i,\sigma}^{\dagger} :=
\sum_{\alpha \in {\cal N}_i}
\lambda_{i,\alpha,\sigma}^{}
c_{i,\alpha,\sigma}^{\dagger}\,.
\end{equation}
They obey the following algebra:
\begin{eqnarray}
\label{eq:3}
&
{[{\Psi}_{i,\sigma}^{\dagger},{\Psi}_{j,\sigma '}^{\dagger}]}_+
=
{[{\Psi}_{i,\sigma}^{},{\Psi}_{j,\sigma '}^{}]}_+
=
0
&
\\
\nonumber
&
{[{\Psi}_{i,\sigma}^{},{\Psi}_{j,\sigma '}^{\dagger}]}_+
=\,\delta_{\sigma,\sigma '}^{}
&
\\
\nonumber
&
\times
\left\{
\begin{array}{ll}
           \sum\limits_{\alpha\in{\cal N}_i}\,
           {|\lambda_{i,\alpha,\sigma}^{}|}_{}^{2}
                                &\quad {\rm for}\quad i=j\\
           \sum\limits_{\alpha,\beta\in{\cal C}_{ij}}
           \delta_{(i\alpha),(j\beta)}^{}
           \lambda_{i,\alpha,\sigma}^{\star}
           \lambda_{j,\beta,\sigma}^{}
                                &\quad {\rm otherwise}\\
\end{array}
\right. \,.
&
\end{eqnarray}
(The Kronecker symbol $\delta_{(i\alpha),(j\beta)}^{}$
assumes the value one if the indices $i\alpha$ and $j\beta$ label
the same site and equals zero otherwise.) The coefficients
$\lambda_{i,\alpha,\sigma}^{}$ define the parameters of the
Hamiltonian which will be constructed below.

The $\Psi$-operators allow to write down a positive
semidefinite\cite{positive} Hamiltonian in a very concise
way:
\begin{equation}
\label{eq:4}
H
=
\sum_{\sigma}
H_{\sigma}
\quad {\rm with} \quad
H_{\sigma}
=
\sum_{i}
{\Psi}_{i,\sigma}^{}
P
{\Psi}_{i,\sigma}^{\dagger}\,.
\end{equation}
However, the physical meaning of this Hamiltonian becomes
more transparent if it is rewritten in terms of the Fermion
$c$-operators by using the relation
\begin{equation}
\label{eq:5}
{\Psi}_{i,\sigma}^{\dagger} P
=
P {\Psi}_{i,\sigma}^{\dagger}
+
\sum_{\alpha\in{\cal U}_i}
\lambda_{i,\alpha,\sigma}^{}
c_{i,\alpha,\sigma}^{\dagger}
n_{i,\alpha,-\sigma}^{}
P
\end{equation}
and the definition of the $\Psi$ operators as defined in
Eq.\ (\ref{eq:2}). The Hamiltonian (\ref{eq:4}) is transformed
into the following form:
\begin{eqnarray}
\label{eq:6}
H_{\sigma} \quad = \quad
              - \,P\,\, \sum_{i} \Big\{
               &{\sum\limits_{\alpha \not= \beta \in {\cal N}_i}}&
                    \lambda_{i,\alpha,\sigma}^{}
                    \lambda_{i,\beta,\sigma}^{\star}\,
                     {c_{i,\alpha,\sigma}^{\dagger}}
                      c_{i,\beta,\sigma}^{}\\
\nonumber
             + \,\, &{\sum\limits_{\alpha\in{\cal N}_i}}&
                   |{\lambda_{i,\alpha,\sigma}^{}}|^{2}\,
                              (n_{i,\alpha,\sigma}^{}-1) \\
\nonumber
             + \,\, &{\sum\limits_{\alpha\in{\cal U}_i}}&
                   |{\lambda_{i,\alpha,\sigma}^{}}|^{2}\,
                              n_{i,\alpha,-\sigma}^{}
                                  \Big\} \,P\,.
\nonumber
\end{eqnarray}
The first term in $H_{\sigma}$ describes particle
hopping within a cell where each site (orbital) is connected
to all others. (Particle transfer between interacting
orbitals and orbitals without interaction is usally called
a hybridization. In that case, the model would describe
e.\,g.\ an Anderson model.) The remaining terms contain trivial
constants and sums over occupation numbers which add up to the
total particle-number operator for the case of the $D$-dimensional
Hubbard model as discussed below. If the sets ${\cal U}_i$ and
${\cal N}_i$ are not identical or if not every site within a cell
is connected to a neighbor cell, the sums over occupation number
operators may add up to an additional field which shifts the
energy of certain sites. Such a model then contains more than one
parameter and the exact ground state as derived below is only
valid on a parameter surface defined by the quantities
$\lambda$. In the following, all $\lambda$-coefficients are
assumed to be real\cite{real}.

\section{The ground state for 2$N$ or more particles}
\label{sec:3}

The models as defined above allow to write down an
exact ground state if the system contains two or more
particles per cell. The ground state is a Gutzwiller
projected Slater determinant:
\begin{equation}
\label{eq:7}
|\Phi_0\rangle
:=
P
\prod_i
{\Psi}_{i,\uparrow}^{\dagger}
{\Psi}_{i,\downarrow}^{\dagger}
|\chi\rangle \,.
\end{equation}
The expression $H |\Phi_0\rangle$ always contains
a factor ${(P{\Psi}_{i,\sigma}^{\dagger})}^{2}$.
Because of the equation
$
{(P
{\Psi}_{i,\sigma}^{\dagger})}^{2}
=0
$,
the state Eq.\ (\ref{eq:7}) is an eigenstate of $H$
(\ref{eq:4}), the corresponding eigenvalue equals zero.

Further, $H$ is positive semi-definite\cite{positive}.
Therefore the state (\ref{eq:7}) belongs to the ground-state
manifold. In Ref.\onlinecite{tasaki2} Tasaki provides a proof of
the uniqueness of the ground state if the state $|\chi\rangle$
is equal to the vacuum state $|0\rangle$ ($2N$ particles).
Larger fillings can be discussed if $|\chi\rangle$ contains
additional particles, for that case it immediately follows
by construction that the state (\ref{eq:7}) is degenerate.

If all sites carry an infinitely strong on-site repulsion and,
e.\,g.\ the $\lambda$-coefficients are set to unity, the state
(\ref{eq:7}) exhibits a structure often called
resonating-valence-bond state.\cite{pauling} In that case the
ground state consists of a linear combination of local singlet
bonds:\cite{tasaki1}
$
P
{\Psi}_{i,\uparrow}^{\dagger}
{\Psi}_{i,\downarrow}^{\dagger}
=
P
\sum_{\alpha\neq\beta}
\big(
c_{i,\alpha,\uparrow}^{\dagger}
c_{i,\beta,\downarrow}^{\dagger}
+
c_{i,\beta,\uparrow}^{\dagger}
c_{i,\alpha,\downarrow}^{\dagger}
\big)
$.
The projector ensures that no two singlet bonds have a
site in common.

\section{Upper bound on the ground-state energy
for 2$N$-1 particles}
\label{sec:4}

Although the ground state is known for two or more particles
per cell, the only analytical result that has been obtained for
particle numbers $N_e$ below the ``magic'' number $N_e=2N$ so
far is the trivial lower bound on the ground state
energy $E_0(N_e)\ge 0$ which is valid for {\em any} number of particles.
[The Hamiltonian (\ref{eq:4}) is non-negative.] The present section
provides an upper bound on the ground state energy of $H$ for one
additional hole, e.\,g.\, a particle number of $N_e=2N-1$
using a variational state.
The quality of variational methods is not very well
controlled, there exists no small parameter, only an
inequality for the energies. Therefore the method requires
some intuition in order to ``hit'' the correct physics by
guessing a good state. The first guess of a variational state,
the state
\begin{equation}
\label{eq:8}
|\varphi_i\rangle
:=
P
{\Psi}_{i,\uparrow}^{\dagger}
\prod_{j,\,\, j\neq i}
{\Psi}_{j,\uparrow}^{\dagger}
{\Psi}_{j,\downarrow}^{\dagger}
|0\rangle \,,
\end{equation}
is led by the idea to leave as much of the $2N$ -particle
physics unchanged as possible. The state is constructed
in a way that most terms of the Hamiltonian generate
zero-states due to the relation
$(P\Psi_{i,\alpha,\sigma}^{\dagger})^2=0$.
{}From
$
H_{\uparrow}
|\varphi_i\rangle
=0
$
and
$
H_{\downarrow}
|\varphi_i\rangle
=
{\Psi}_{i,\downarrow}^{}
|\Phi_0\rangle
$
($|\Phi_0\rangle$ denotes the ground state of the system
with $2N$ particles of the previous section) one obtains the
diagonal matrix
\begin{equation}
\label{eq:9}
\langle \varphi_j|
H
|\varphi_i\rangle
=
\delta_{i,j}^{}
\langle\Phi_0|\Phi_0\rangle\, .
\end{equation}
If the norm of the states $|\varphi_i\rangle$
and $|\Phi_0\rangle$ were known, the expectation
value of $H$ could be evaluated. Unfortunately, this is
not possible in general. However, the following
orthogonal and normalized [see Eq.\ (\ref{eq:9})]
trial states
\begin{equation}
\label{eq:10}
|\chi_i\rangle
:=
\frac
{\sqrt{H}}
{\sqrt{\langle\Phi_0|\Phi_0\rangle}}
|\varphi_i\rangle
\end{equation}
allow us to calculate the desired expectation value.
(The positive square root of $H$ is well defined because
$H \geq 0$.) Surprisingly, the Hamiltonian matrix
elements $\langle\chi_j|H|\chi_i\rangle$ of
the $(2N-1)$ -particle system can be expressed in
terms of occupation-number expectation-values of
the $2N$-particle system in the following way:
\begin{eqnarray}
\label{eq:11}
&
\langle
\chi_j|
H
|\chi_i\rangle
=
\delta_{i,j}^{}
\Big(
\sum\limits_{\alpha\in {\cal N}_i}
{\lambda_{i,\alpha,\downarrow}^{}}^{2}
&
\\
\nonumber
&
-
\sum\limits_{\alpha\in {\cal U}_i}
{\lambda_{i,\alpha,\downarrow}^{}}^{2}
\langle n_{i,\alpha,\uparrow}^{}\rangle
\Big)
&
\\
\nonumber
&
+
(1-\delta_{i,j}^{})
\Big(
\sum\limits_{\alpha,\beta\in {\cal C}_{ij}}
\delta_{(i\alpha),(j\beta)}^{}
\lambda_{i,\alpha,\downarrow}^{}
\lambda_{j,\beta,\downarrow}^{}
&
\\
\nonumber
&
-
\sum\limits_{\alpha,\beta\in \{{\cal C}_{ij}\cap{\cal U}_i \}}
\delta_{(i\alpha),(j\beta)}^{}
\lambda_{i,\alpha,\downarrow}^{}
\lambda_{j,\beta,\downarrow}^{}
\langle n_{i,\alpha,\uparrow}^{}\rangle
\Big)\,,
&
\end{eqnarray}
where
\begin{equation}
\label{eq:12}
\langle n_{i,\alpha,\sigma}^{}\rangle
=\frac{{\displaystyle \langle\Phi_0|n_{i,\alpha,\sigma}^{}|\Phi_0\rangle}}
                {{\displaystyle \langle\Phi_0|\Phi_0\rangle}}
\,.
\end{equation}
In the following section, the upper bound Eq.\ (\ref{eq:11})
on the ground-state energy is discussed for two examples
which allow us to calculate the expectation value
$\langle n_{i,\alpha,\sigma}^{}\rangle$.

\section{Results}
\label{sec:5}

\subsection{$D$-dimensional Hubbard model}
\label{sec:5.1}

The first example, where the upper bound on the ground-state
energy of the $(2N-1)$ -particle system may be evaluated is
the Hubbard model as introduced in Ref.\onlinecite{brandt-giesekus},
where all $\lambda$'s are set to unity and the spatial dimension
is larger than one. For the three-dimensional case, the cells have
the topology of octahedra with $2D=6$ sites. All sites carry
interaction, therefore the sets ${\cal U}_i$ and ${\cal N}_i$ are
identical. Further, every site in ${\cal N}_i$ is shared by exactly
one neighboring cell, forming a $D$ dimensional (hyper)cubic lattice
of cells. Periodic boundary conditions are assumed. The corresponding
lattice is visualized in Fig.\ \ref{fig:1}a. The Hamiltonian of
this system reads
\begin{eqnarray}
\label{eq:13}
H
&=&
\sum_{i,\sigma}
\Psi_{i,\sigma}^{}
P
\Psi_{i,\sigma}^{\dagger}\\
\nonumber
&=&
P
\sum_{i,\sigma}
\Big(
-
\sum_{{\scriptstyle \alpha,\beta=1}
 \atop{\scriptstyle \alpha\neq\beta}}^{2D}
c_{i,\alpha,\sigma}^{\dagger}
c_{i,\beta,\sigma}^{}
\Big)
P
+
4\,
P
\big(
ND - \hat{N}_e
\big)\,.
\end{eqnarray}
The last term in (\ref{eq:13}) contains a trivial constant $4ND$
and the total electron-number operator $\hat{N}_e$ (with eigenvalue
$N_e$) which is a conserved quantity.

With the Fourier transform
\begin{equation}
\label{eq:14}
|\chi_k\rangle
=
\frac{1}{\sqrt{N}}
\sum_{j}
e^{ikj}
|\chi_j\rangle
\end{equation}
and $\langle n_{i,\alpha,\sigma}^{}\rangle=D^{-1}$ (the occupation
probabilities for all sites are equal for symmetry reasons) one
obtains an upper bound\cite{dispersion1} on the $(2N-1)$ -particle
ground state energy
\begin{equation}
\label{eq:15}
\langle\chi_k|H|\chi_k\rangle
=2(1-\frac{1}{D})(D+\sum_{\mu=1}^{D}\cos k_{\mu})
\end{equation}

The minimum
$
E_{>}(2N-1)=\min\,\langle\chi_k| H|\chi_k\rangle
$
with respect to the $k$-vector is obvious: For
$(k_1 , k_2 , \ldots , k_D)=(\pi,\pi,\ldots,\pi)$
the right hand side of (\ref{eq:15}) becomes zero.
However, this $k$-vector has to be {\em excluded} for this
particular model, because it can be shown\cite{brandt-giesekus}
(c.\,f.\, App.\ \ref{app:1}) that the state $|\Phi_0\rangle$
(\ref{eq:7}) vanishes if $\vec{k}=(\pi,\ldots,\pi)$ is allowed.
That particular $k$-vector can be excluded by the restriction to
systems where one of the $D$ system dimensions contains
an odd number of cells. Then the $k$-vector component
corresponding to this particular direction cannot equal
$\pi$.

This result indicates that the state $|\chi_k\rangle$ becomes
an asymptotically exact $(2N-1)$ -particle ground state of
$H$ for the thermodynamic limit. The upper bound on the
ground state energy vanishes as
\begin{equation}
\label{eq:16}
E_{>}(2N-1)= O(N^{-\frac{2}{D}})
\end{equation}
and asymptotically approaches the lower bound $E_{<}(2N-1)=0$.
This result is even relevant for $D=2$, where the asymptotically
exact state describes one hole in a half filled RVB-background.
The absence of ferromagnetism does not contradict Nagaoka's
theorem\cite{nagaoka} because the present lattice is not
bipartite. (For a bipartite lattice, the state would be a
ferromagnet, i.\,e.\ the total spin would assume its maximum value.)
Unfortunately, no rigorous statements on the uniqueness of this
ground state can be made.

Further, it can be seen that for $D>2$ the chemical potential
of the system for particle numbers of $N_e=2N-1,\ldots,ND$ remains
constant. The non-interacting tight-binding analogue exhibits quite
similar behavior: The band structure of this particular lattice
consists of one cosine band with
$
\epsilon(k)=2-2(D+\sum_{\mu}\cos k_{\mu})
$
and $D-1$ flat bands at the upper band edge of the extended band.
However, the ``bandwidth'' of the interacting system is reduced by
a factor $(1-1/D)$, which may be interpreted as an increased mass.
This result sheds new light on the conductivity in the system. It
might be that the many-particle state describes a dispersive
delocalized hole which may contribute to conduction. Unfortunately,
no results on hole {\em densities} have been obtained and
the dispersive state is not exact except in the vicinity of
$\vec{k}=(\pi,\ldots,\pi)$.

\subsection{Linear Hubbard chain}
\label{sec:5.2}

As a second example, we study a linear chain with $N$
cells and $4$ sites per cell ${\cal N}_i$ ($3$ sites
per {\em unit cell}). The topology of this chain is
illustrated in Fig.\ \ref{fig:1}b; it may be interpreted
as a chain of connected tetrahedra. Again, periodic
boundary conditions and the case of infinite
on-site repulsion on every site are assumed, and thus
the sets ${\cal N}_i$ and ${\cal U}_i$ are identical.
The cells $i$ and $i+1$ are connected by one site
(backbone site): the sites with indices
$(i,\alpha=1)$ and $(i+1,\beta=4)$ are identical.
All $\lambda$'s are set to unity. The Hamiltonian of
this particular chain reads
\begin{eqnarray}
\label{eq:17}
H \quad &=& \sum_{i,\sigma} {\Psi}_{i,\sigma}^{} P
                               {\Psi}_{i,\sigma}^{\dagger}\\
\nonumber
              &=&  \quad P \, \Big(
              - \,\,  \sum_{i,\sigma}
                    \sum_{{\alpha,\beta=1}
                          \atop
                          {\alpha \not= \beta}}^{4}
                             {c_{i,\alpha,\sigma}^{\dagger}}
                              c_{i,\beta,\sigma}^{}
             - \,\, 2 \, \sum_{i,\sigma} n_{i,1,\sigma}^{}
                                  \Big) \,P\\
\nonumber
  &{}&\quad\quad\quad +\,\,P\,\left( 8N - 2 \hat{N}_e \right)\,.
\end{eqnarray}
In addition to the hopping terms, one obtains a local field
which shifts the energy of the backbone sites ($\alpha=1$).

For this model, the upper bound Eq.\ (\ref{eq:11})
reduces to the simple expression\cite{dispersion2}
\begin{equation}
\label{eq:18}
\langle\chi_k|H|\chi_k\rangle
=
3 -
\langle n_{i,\alpha=1,\uparrow}^{}\rangle
+
2
(1-\langle n_{i,\alpha=1,\uparrow}^{}\rangle)
\cos k
\end{equation}

The right hand side of Eq.\ (\ref{eq:18}) assumes a minimum
for $k=\pi$ (which is allowed for this model). This leads
to the following upper bound on the ground state energy:
\begin{equation}
\label{eq:19}
E_{>}(2N-1)=1+\langle n_{i,\alpha=1,\uparrow}^{}\rangle \,.
\end{equation}
The occupation probability on the backbone sites may be
calculated by a transfer-matrix technique for the
thermodynamic limit. This calculation is briefly outlined
in appendix \ref{app:2} and yields
\begin{equation}
\label{eq:20}
\lim_{N\to\infty} E_{>}(2N-1)=
\frac{1}{2}(3-\frac{1}{\sqrt{13}})\approx 1.361325
\end{equation}
In contrast to the previous example this result indicates a
gap in the $(2N-1)$ -particle spectrum. Equation (\ref{eq:20})
estimates this gap from above. Therefore the upper bound is compared
to numerically obtained exact results as described in
Ref.\onlinecite{giesekus}.

The numerical results have been obtained by the conjugate-gradient
algorithm.\cite{bradbury,nightingale} The energies are listed in
Tab.\ \ref{tab:1}. For chain lengths up to four unit cells, the
ground state is calculated without applying symmetry restrictions.
In order to check for ground-state degeneracy, the determination of
the lowest eigenstate is repeated in the subspace orthogonal to the
ground state. For an even number of unit cells, the energies of
these orthogonal eigenstates do not coincide with the ground-state
energy, therefore the ground states for even chain length are
non-degenerate. (Chains with an odd number of unit cells are
degenerate because of reflection symmetry which induces degeneracy
with respect to $\pm k$.)

As a second step, the same calculations have been performed
assuming site-interchange symmetry as decribed in
Ref.\onlinecite{giesekus} and translational invariance with
$k=\pi$. The energies of the symmetrized states coincide
with the previously calculated non-degenerate ground-state
energies, therefore the symmetry of the ground state is uniquely
determined.

The chain with six cells is numerically treated by application
of site interchange symmetry only and the result for $N=8$ is obtained
assuming translational invariance in addition to site-interchange
symmetry.

The numerically obtained exact ground-state energies $E_0 (2N-1)$ are
shown in Tab.\ \ref{tab:1}. The upper bound Eq.\ (\ref{eq:20}) of
the infinite system deviates from the numerical result for $N=8$
by $\approx 13\%$. This agreement is not very close. However, the
symmetries of the variational state and the numerically exact state
coincide. Better quantitative agreement may be obtained by the
expectation value of $H$ with respect to the state
\begin{equation}
\label{eq:21}
|\varphi_k\rangle
=
\frac{1}{\sqrt{N}}
\sum_{j}
e^{ikj}
|\varphi_j\rangle\,,
\end{equation}
[The state $|\varphi_i\rangle$ is defined by Eq.\ (\ref{eq:8}).], which
may be evaluated numerically. For a chain with $N=4$ one obtains
$\langle\varphi_{k=\pi}|H|\varphi_{k=\pi}\rangle=1.2555$.
This result deviates from the numerically exact result by
$\approx 4\%$ only. The state $|\chi_k\rangle$ is proportional
to $\sqrt{H}|\varphi_k\rangle$, therefore a difference in the
expectation values $\langle\varphi_{k=\pi}|H|\varphi_{k=\pi}\rangle$
and $\langle\chi_{k=\pi}|H|\chi_{k=\pi}\rangle$ simply reflects the
fact that $|\chi_{k=\pi}\rangle$ (and hence $|\varphi_{k=\pi}\rangle$)
is not an eigenstate. The operator $\sqrt{H}$ ``amplifies'' the
deviation from the exact ground state.

The single-particle band-structure of this particular chain
is illustrated in Fig.\ \ref{fig:2} and consists of two extended
bands, one broad band separated from a narrower one by a gap of
the size of one energy unit. The highest energy band collapses to
one dispersionless flat band which is again separated by a small
gap. For $2N$ particles the non-interacting system would have a
completely filled lower valence band describing an insulator. It
is obvious that the interacting system shows different behavior
here: In the framework of the one-particle language, one would
expect a gap if the particle number is changed from $N_e=2N$ to
$N_e=2N+1$, however, for that case the chemical potential of the
interacting system remains constant. In contrast to the one-particle
picture, the chemical potential of the interacting system changes
if the particle number is changed from $N_e=2N-1$ to $N_e=2N$. This
contradiction illustrates the breakdown of one-particle physics in
this particular model.

\section{Summary}
\label{sec:6}

For a class of models describing strongly interacting Fermions
it is possible to write down the exact ground state if the
particle number equals 2 per unit cell or more. The particle
number regime below this threshhold has not been accessible
by rigorous methods so far. We present a tractable variational state
for systems with one additional hole. The corresponding expectation
value of the Hamiltonian approaches a lower bound for the case of a
particular $D$-dimensional Hubbard model in the thermodynamic limit.
We conclude that the presented variational state approaches the ground
state or an eigenstate which is degenerate to the ground state if the
ground state is not unique.

For a second model, a special Hubbard chain, the analytical
expectation value deviates significantly from the numerically
calculated ground state energy. Although it is shown that the
numerical ground state and the variational state exhibit the
same symmetries, it remains unclear, how well the ground-state
physics of the chain is described by our variational ansatz.

\acknowledgments

We gratefully acknowledge numerous helpful discussions with
J.\,Stolze. This work has been supported by the Deutsche
Forschungsgemeinschaft (DFG), Project Br\,434/6-2.

\appendix
\section{}
\label{app:1}

This appendix provides a proof that the state $|\Phi_0\rangle$
[Eq.\ (\ref{eq:7})] of the $2N$- particle $D$- dimensional
system vanishes, if the system extends an even number of unit
cells in each direction.

The state and the Hamiltonian are expressed in terms of creation
(annihilation) operators $\Psi_{i,\sigma}^{\dagger}$
($\Psi_{i,\sigma}^{}$) referring to non-orthogonal one-particle
states. However, they must be linear independent, otherwise the
corresponding Slater determinant vanishes.

There is no problem for models with non-equivalent sites,
however, the $D$-dimensional Hubbard model requires some
care. Consider the Slater determinant
\begin{equation}
\label{eq:a1}
|\Phi\rangle=
\prod_{i}
\Psi_{i,\uparrow}^{\dagger}
\Psi_{i,\downarrow}^{\dagger}
|0\rangle\,.
\end{equation}
The norm of $|\Phi\rangle$ can be
expressed in terms of a determinant
\begin{equation}
\label{eq:a2}
\sqrt{\langle\Phi|\Phi\rangle}
=
\det {\rm\bf M}
\end{equation}
The matrix ${\rm\bf M}$ has the elements
\begin{equation}
\label{eq:a3}
M_{ij}=2D\delta_{ij} + \delta_{<ij>} \,,
\end{equation}
where $\delta_{<ij>}$ equals one if $i$ and $j$ are
nearest neighbors and zero otherwise. The structure
of ${\rm\bf M}$ is similar to a $D$-dimensional
hypercubic tight-binding problem with periodic boundary
conditions.

The determinant is equal to the product over all
eigenvalues of the above tridiagonal matrix. The
eigenvalues are given by the familiar cosine bands
$2(D+\sum_{\nu=1}^{D}\cos k_{\nu})$. If all system
length contain an even number of unit cells, one of
the eigenvalues vanishes, because one of all possible
$k$- vectors equals $(\pi,\pi,\pi,\ldots,\pi)$.
Therefore, the state $|\Phi\rangle = 0$ and the
projected state $|\Phi_0\rangle=P|\Phi\rangle$
vanishes as well. The condition that at least one
system length must be odd is a {\em necessary} condition
for this model.

\section{}
\label{app:2}

We briefly recall the transfer matrix calculation
of the occupation probability for certain sites of
the Hubbard chain. This calculation requires open
boundary conditions. However, the influence of the
boundary conditions decreases with increasing system
size.

The ground state of this chain can be decomposed into
a sum over all possible configurations of $N$ singlet
bonds (dimers). Every component of the ground state has
exactly one dimer in each cell. Let the singlets
(dimers) be written as
$
b_{\alpha\beta}^{\dagger}|0\rangle=\frac{1}{2}
(c_{i,\alpha,\uparrow}^{\dagger}
c_{i,\beta,\downarrow}^{\dagger}
+
c_{i,\beta,\uparrow}^{\dagger}
c_{i,\alpha,\downarrow}^{\dagger})|0\rangle
$.

For every cell there exist six possible contributions:
\begin{eqnarray}
\nonumber
|1\rangle = b_{12}^{\dagger}|0\rangle\quad
|2\rangle &=& b_{13}^{\dagger}|0\rangle\quad
|3\rangle = b_{14}^{\dagger}|0\rangle\\
\label{eq:b1}
\\
\nonumber
|4\rangle = b_{42}^{\dagger}|0\rangle\quad
|5\rangle &=& b_{43}^{\dagger}|0\rangle\quad
|6\rangle = b_{23}^{\dagger}|0\rangle
\end{eqnarray}
The interaction excludes any double occupancy. Therefore
the indices $\alpha$ and $\beta$ have to be different
within a cell. This condition is not sufficient, because
double occupancy on the backbone sites could occur due
to certain configurations of neighbor cells: If e.\,g.\ cell
$i$ contains the singlet $|1\rangle$ and cell $i+1$ the
singlets $|3\rangle$, $|4\rangle$, or $|5\rangle$ the
connection site would be occupied by two particles. These
neighbor configurations have to be excluded. The restriction
to allowed neighbor configurations is coded by the transfer
matrix:
\begin{equation}
\label{eq:b2}
T=\left(
\begin{array}{cccccc}
1&1&0&0&0&1\\
1&1&0&0&0&1\\
1&1&0&0&0&1\\
1&1&1&1&1&1\\
1&1&1&1&1&1\\
1&1&1&1&1&1\\
\end{array}
\right)\,.
\end{equation}
With these definitions, the norm of the ground state
$\langle\Phi_0|\Phi_0\rangle$ for a chain with $N$
cells may be written as
\begin{equation}
\label{eq:b3}
\langle\Phi_0|\Phi_0\rangle=
\sum_{i,j} (T^{N-1})_{ij}\,.
\end{equation}
The occupation of one $\alpha=2$-site well within
the chain (the cell index $m$ should be $\approx 1/2 N$)
is given by the expression
\begin{equation}
\label{eq:b4}
\langle (n_{i,2,\uparrow}^{} + n_{i,2,\downarrow}^{})\rangle =
\frac{1}{\langle\Phi_0|\Phi_0\rangle}
\sum_{i,j} \sum_{l=1,4,6} (T^{m})_{il}(T^{N-m-1})_{lj}
\end{equation}
which may be simplified by using a spectral representation
of the matrix $T$.
The eigenvalues of $T$ are
$\epsilon_+=\frac{1}{2}(5+\sqrt{13})$,
$\epsilon_-=\frac{1}{2}(5-\sqrt{13})$ and a fourfold
degenerate eigenvalue $\epsilon=0$. The right and left
eigenstates may be calculated by
$T|r_{\pm}\rangle=\epsilon_{\pm}|r_{\pm}\rangle$
and
$\langle l_{\pm}|T=\epsilon_{\pm}\langle l_{\pm}|$.
They are mutually orthogonal and have to be normalized
according to $\langle l_{\pm}|r_{\pm}\rangle=1$. Then the
matrix $T^{N}$ reads
\begin{equation}
\label{eq:b5}
T^{N}\,\,=\,\,|r_+ \rangle (\epsilon_+)^{N} \langle l_+|
       \quad+\quad
              |r_- \rangle (\epsilon_-)^{N} \langle l_-|
\end{equation}
As usual, the contribution of the largest eigenvalue dominates
the results with growing system size. In the thermodynamic
limit the total occupation of the $\alpha=2$-sites becomes
\begin{equation}
\label{eq:b6}
\langle (n_{i,2,\uparrow}^{} + n_{i,2,\downarrow}^{})\rangle =
\sum_{i=1,4,6}
\langle l_+|i\rangle\langle i| r_+ \rangle
=\frac{1}{2}(1+\frac{1}{\sqrt{13}})
\end{equation}
The backbone occupation may be calculated analogously or simply
by applying
$\sum_{\sigma}\sum_{\alpha=1}^{3}
\langle n_{i,\alpha,\sigma}^{}\rangle = 2$.

%
%

%
%

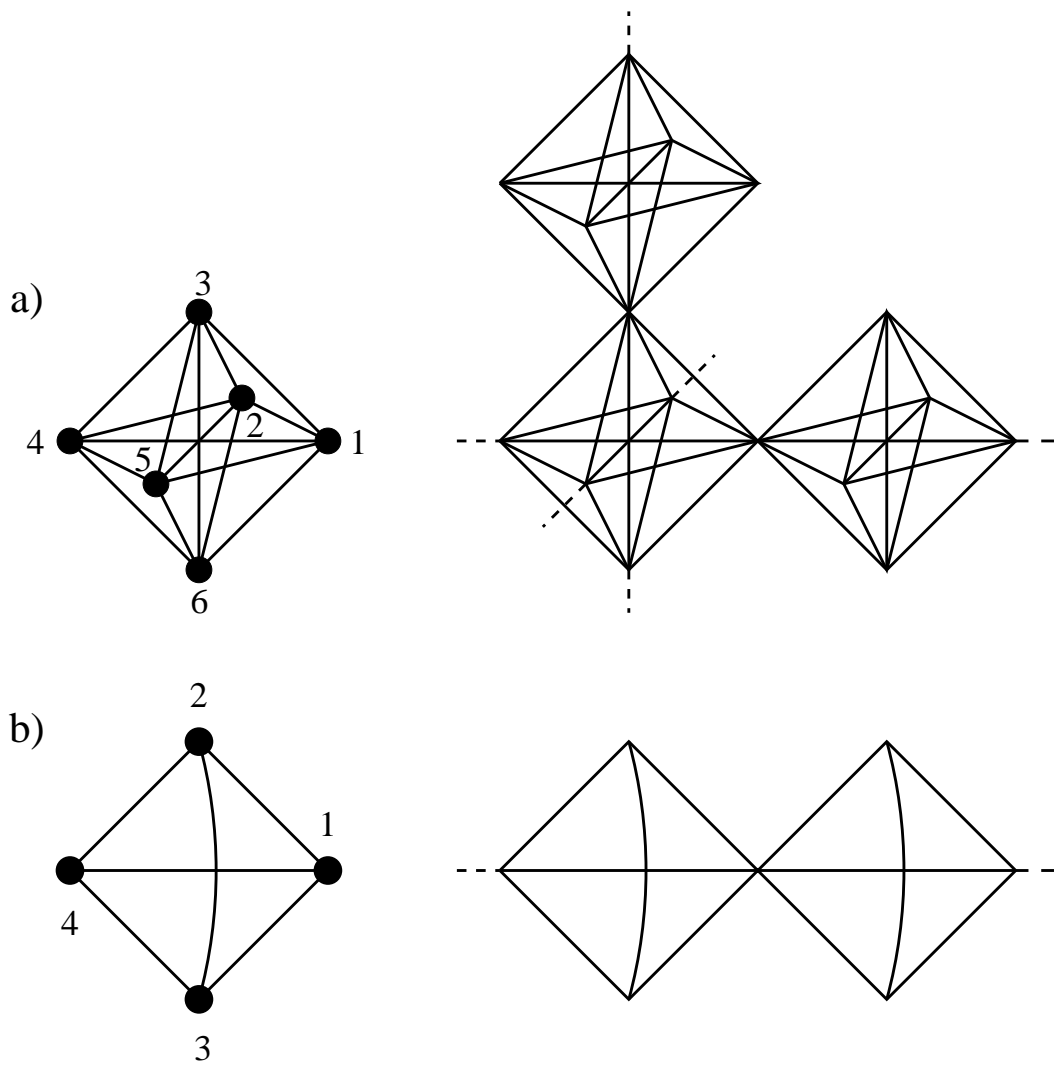
\begin{figure}
\caption{
Two examples of lattice structures discussed here. The
left column shows the cells which are connected to a
lattice (chain). Filled circles denote sites with infinite
on-site repulsion, solid lines illustrate the hopping.
a) Lattice of the hypercubic model for three dimensions.
The sites within a unit cell are labeled by an index
$\alpha=1,\ldots,2D$. The lattice consists of
connected octahedra located at the vertices of a hypercubic
lattice.
b) One-dimensional Hubbard chain with four sites
$\alpha=1,\ldots,4$ per cell.
}
\label{fig:1}
\end{figure}
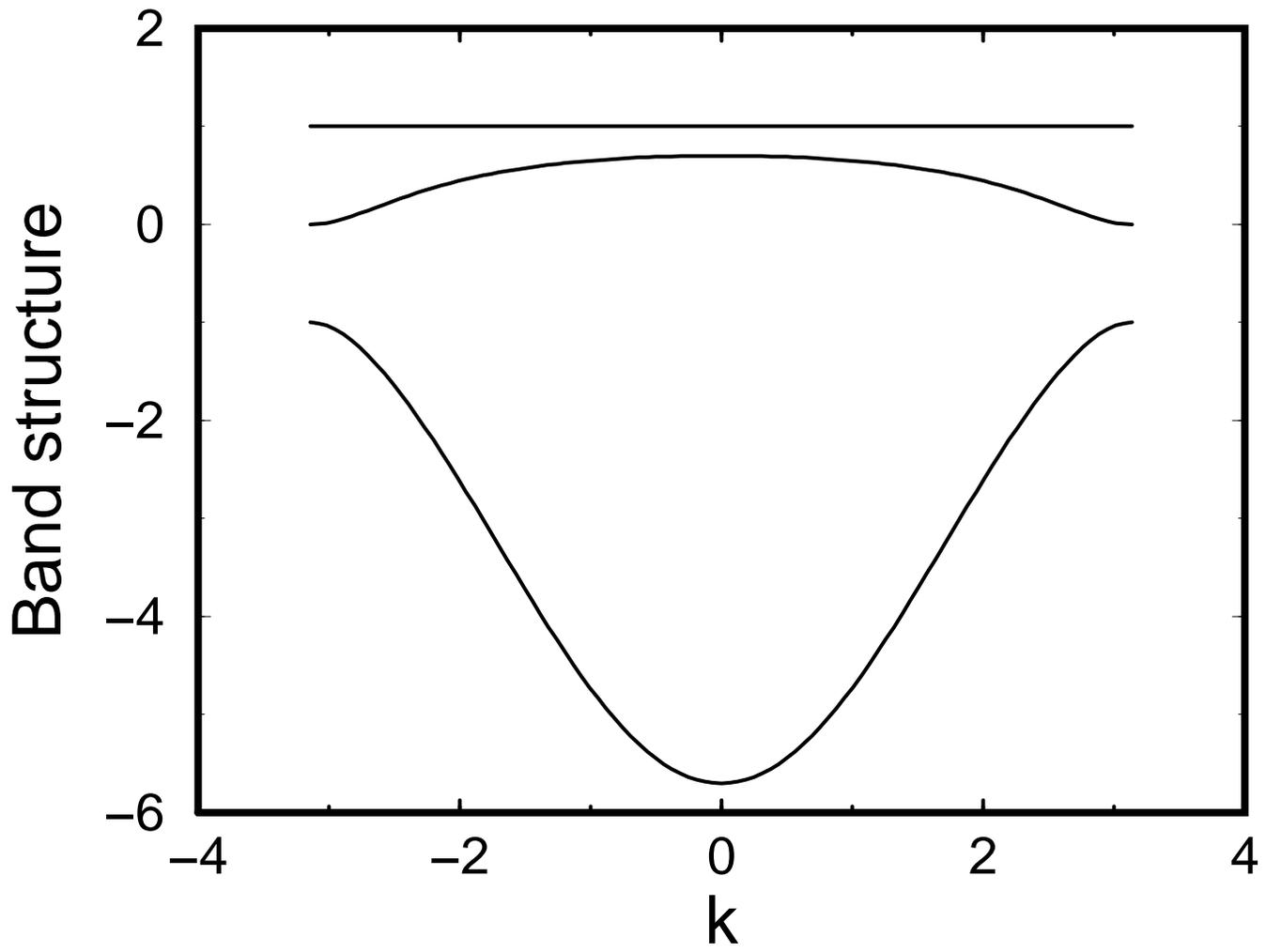
\begin{figure}
\caption{
Band structure of the non-interacting chain. The spectrum shows
two dispersive bands separated by a gap of one energy unit. The
third band collapses to a flat band leading to a $\delta$- peak
in the corresponding density of states.
}
\label{fig:2}
\end{figure}
%

\begin{table}
\caption{Numerically calculated ground-state energies for
systems with $N$ unit cells and $2N-1$ particles. The states
are unique and correspond to $k=\pi$.}
\begin{tabular}{cd}
$N$&ground-state energy\\
\hline
2 & 1.2275046 \\
4 & 1.2137261 \\
6 & 1.2099690 \\
8 & 1.2095401 \\
\end{tabular}

\label{tab:1}
\end{table}


\begin{references}
\bibitem{positive} Because of $P^2=P$ (projector) the
                   Hamiltonian (\ref{eq:4}) has a structure
                   $\Sigma AA^{\dagger}$.
                   Any expectation value obeys
                   $\langle\varphi|H|\varphi\rangle
                   = \Sigma
                   {||A^{\dagger}|\varphi\rangle||}^2
                   \geq 0$.
\bibitem{real}It is possible to construct a model with periodic
              boundary conditions, where not all phases of the
              $\lambda$-coefficients can be removed by a gauge
              transformation.\cite{tasaki2} The implications of
              this fact are disregarded here.
\bibitem{dispersion1}Equation (\ref{eq:15}) provides an upper
                     bound on the one-hole dispersion relation
                     which is gapless for this model.
\end{references}
\end{document}